\documentclass[twocolumn,9pt]{article} 

\usepackage[square,numbers,sort&compress,comma]{natbib}
\usepackage{multirow}
\usepackage{amsmath}
\usepackage{amssymb}
\usepackage{caption}
\usepackage{graphicx}
\usepackage{latexsym}
\usepackage{times}
\usepackage[pagewise]{lineno}
\usepackage{hyperref}
\usepackage{comment}
\usepackage{xcolor}
\topmargin - 12pt 
\renewenvironment{abstract}%
{
	\small
	{\bfseries \abstractname}
	\par
	\vspace{10pt}
}

\renewcommand\abstractname{Abstract}

\newcommand{\nomenclature}
[1]
{
	\bgroup
	\flushleft
	\small\bf
	#1
	\par
	\egroup
}

\renewcommand{\section}
[1]
{
	\bgroup
	\flushleft
	\small\bf
	\refstepcounter{section}
	\arabic{section}. #1
	\par
	\egroup
}

\renewcommand{\subsection}
[1]
{
	\bgroup
	\flushleft
	\small\em
	\refstepcounter{subsection}
	\arabic{section}.
	\arabic{subsection}. #1
	\par
	\egroup
}

\renewcommand{\subsubsection}
[1]
{
	\bgroup
	\flushleft
	\small\em
	\refstepcounter{subsubsection}
	\arabic{section}.
	\arabic{subsection}.
	\arabic{subsubsection}. #1
	\par
	\egroup
}

\newcommand{\acknowledgement}
[1]
{
	\bgroup
	\flushleft
	\small\bf
	#1
	\par
	\egroup
}

\newcommand{\sectionbib}
[1]
{
	\bgroup
	\flushleft
	\small\bf
	#1
	\par
	\egroup
}

\begin{document}
	
	
	
	\small
	\baselineskip 10pt
	
	\setcounter{page}{1}
	\title{\LARGE \bf Burning limits and the role of small scale turbulence\\ 
		in high Karlovitz number premixed flames}
	
	\author{{\large Praful Kumar$^{a,*}$, \"{O}mer L. G\"{u}lder$^{a}$}\\[10pt]
		{\footnotesize \em $^a$Institute for Aerospace Studies, University of Toronto}\\[-5pt]
		\footnotesize \em 4925 Dufferin Street, Toronto, ON, Canada M3H 5T6}
	
	\date{}  
	
	\twocolumn[\begin{@twocolumnfalse}
		\maketitle
		\rule{\textwidth}{0.5pt}
		\vspace{-5pt}
		
		\begin{abstract} 
			Premixed flames exhibit a progressive enhancement in burning rates with turbulence, with the ultimate limits governed by the onset of local extinctions. Piloted premixed burners have enabled access to turbulence far exceeding the theoretically predicted extinction limits, with the measured flame speeds still improving. This study explores the role of small scale turbulence on the burning rates of premixed flames using a piloted high Karlovitz number burner. Premixed CH$_4$-air mixtures with equivalence ratios ($\phi$) from 0.7--1.0, stabilized using a $\phi$-matched pilot flame, were subjected to increasing rms fluctuating velocities, $u'$, up to 36.0 m/s. The experimental setup enables a progressive increase in turbulent kinetic energy (TKE) at smaller scales, with the Kolmogorov scales reaching approximately 4 $\mu$m under the highest turbulence condition. Global consumption flame speeds, $S_\mathrm{T,GC}$, derived from the binarized Planar Laser-Induced Fluorescence (PLIF) imaging of OH radicals, scale with turbulent diffusivity, $\mathcal{D}_\mathrm{T}$. Further increases in $u'$, however, lead to a maximum and an eventual decline in flame speeds. The measured flame speeds exhibit a common scaling with the measured upstream $u'$. Assessments of the flame structure using CH-PLIF images further reveal progressive straining and extinction events with increasing $u'$ despite flames maintaining their flamelet-like appearance, suggesting that the observed limits derive from the increasing extinction propensity across the flame front. Finally, these findings are incorporated into a scaling relation accounting for the role of straining and extinctions in the observed flame-speed trends.
		\end{abstract}
		
		\vspace{10pt}
		
		{\bf Novelty and significance statement}
		
		\vspace{10pt}
		This study explores the role of small scale turbulence on premixed flames. For the first time, the presented measurements show the limits of turbulence-induced enhancement in burning rates realized using piloted premixed burners. The normalized flame speeds further reveal a consistent scaling with the upstream turbulence despite different mixture compositions, providing evidence that chemistry is superseded by transport in these high Karlovitz number regimes. CH-PLIF imaging highlights progressive straining and extinction events with increasing turbulence, while the reacting flame front remains localized within laminar-like structures, an observation not previously reported in piloted Bunsen flames.
		\vspace{5pt}
		
		\parbox{1.0\textwidth}{\footnotesize {\em Keywords:} Turbulent premixed flames; Extreme turbulence; High Karlovitz number; Flame structure; Turbulent flame speeds;}
		\rule{\textwidth}{0.5pt}
		*Corresponding author.
		\vspace{5pt}
	\end{@twocolumnfalse}] 
	
	\section{Introduction\label{sec:introduction}} \addvspace{10pt}
	When acted upon by turbulence, premixed flames exhibit a progressive increase in their burning rates. These interactions were categorized into coarse-scale (DH1) and fine-scale (DH2) interactions, with coarser scales wrinkling the flame-front and the small scales enhancing local mixing \cite{damkohler1947effect}. A proportionality of the form $S_\mathrm{T}/S_\mathrm{L} \propto {\mathcal{D}_\mathrm{T}} = {(u'\Lambda)}^{0.5}$ was presumed for the role of small scale turbulence, where $S_\mathrm{T}$ and $S_\mathrm{L}$ are the turbulent and laminar flame speeds, $\mathcal{D}_\mathrm{T}$ denotes turbulent diffusivity, and $u'$ and $\Lambda$ are the rms fluctuating velocity and integral length scales of turbulence respectively. With an increase in $u'$, the measured flame speeds show a deviation from linearity, marking a transition from coarse-scales turbulence effects towards small scale mixing \cite{yuen2010dynamics,wabel2017turbulent}. Consequently, flame wrinkling effects, which are critical at low $u'$ \cite{filatyev2005,tamadonfar2015effects}, saturate with an increase in $u'$ \cite{gulder2007fsd,yuen2013turbulent,wabel2017turbulent}.
	
	Using large piloted burners, premixed flames have been subjected to progressively larger $u'$, with the measured flame speeds following a monotonic $\mathcal{D}_\mathrm{T}$ scaling. The proposed limits of burning \cite{Peters_2000} suggest that flames should extinguish when turbulence timescales exceed chemical timescales, with Karlovitz number, Ka = 100 marking the onset of the broken reaction regime, where Ka = $(u'/S_\mathrm{L})^{1.5}(\Lambda/\delta_\mathrm{L})^{-0.5}$. This does not appear to hold true in measurements and flames are able to exist well beyond their expected limits. Similarly, depending on the experimental design and reactant compositions, different observations of the underlying flame structures have been reported \cite{yuen2013turbulent,tamadonfar2015effects,skiba2018part1,zhou2015visualization,zhou2015distributed,MOHAMMADNEJAD2020285}. Distributed burning has been shown to exist in turbulent jets \cite{wang2019structure,shi2024assessing} while similar findings have not been confirmed in piloted Bunsen flames where the reaction layers exhibit limited broadening, with thicknesses not exceeding twice those of the corresponding laminar flames \cite{skiba2018part1,MOHAMMADNEJAD2020285}.

At low $u'$, thermodiffusive instabilities suppress the effects of turbulence in hydrogenated CH$_4$-air flames \cite{chaib2026normalization}, with the measured flame speeds showing agreement with the flame morphological parameters when scaled with freely propagating laminar flame parameters \cite{howarth2024thermal}. These preferential diffusion effects become progressively limited with increasing $u'$ \cite{aspden2011turbulence,shi2024assessing,shi2024internal,russell2025turbulence} while the heat release becomes spatially diffuse. These findings do not translate symmetrically to CH$_4$–air mixtures, for which the corresponding effects remain weak even at Karlovitz numbers exceeding 8000 \cite{aspden_day_bell_2019}. The role of density ratios in explaining this transition discrepancy has been hypothesized. Numerical studies of thermonuclear flames in Type 1a star systems suggest that distributed flame structures are convected akin to laminar flamelets by the larger turbulent structures, leading to an overall increase in burning rates \cite{aspden2008turbulence,aspden2010distributed}. Similarly, in expanding spherical flames, $\mathcal{D}_\mathrm{T}$ scaling holds across reactant mixtures when scaled with Markstein diffusivity \cite{chaudhuri2012flame,chaudhuri2013scaling} with preferential diffusion effects remaining relevant \cite{li2025experimental}. Furthermore, these preferential diffusion effects are conditional on the reaction marker \cite{mohammadnejad2021contributions,wang2025experimental,chaib2024experimental} as well as operating pressures \cite{rieth2022enhanced}. 

The findings of these studies and their implications have been the subject of several reviews \cite{DRISCOLL2020,STEINBERG2021100900,hochgreb2023fast,kheirkhah2022revisit} and isolating these effects on the regime diagram \cite{Peters_2000} remains challenging. An active role of the turbulence spectra on the structure of premixed flames has been proposed \cite{ronney,GULDER20071369,nivarti_2019}. Similarly, NH$_3$ and H$_2$ flames and their blends have been extensively discussed as promising carbon-free energy carriers \cite{kobayashi2019science,awad2023characteristics}, with operations in ultra-lean conditions extending these flames towards high Karlovitz number regimes. The nature of premixed flames in the limits of large turbulence (Ka$>100$), and the limits to their burning, or lack thereof, have not yet been fully realized in experiments.

This study reports measurements of turbulent flame speeds and flame structure obtained in a high Karlovitz number burner \cite{kumar2025characteristics}. Premixed CH$_4$-air flames with equivalence ratios, $\phi$, ranging from 0.7 to 1.0, stabilized using a $\phi$-matched pilot flame are studied up to a Karlovitz number of 950 and their global consumption flame speeds, $S_\mathrm{T,GC}$ \cite{DRISCOLL200891}, are reported. The observed flame speed trends are further examined using qualitative assessments of the flame structure conducted using CH-PLIF. Finally, based on the experimental observations, a modified turbulent flame speed scaling that accounts for the influence of small scale turbulence is proposed.
	\section{Experimental methodology}
	\label{section2}
	\begin{figure}[h!]
		
		\includegraphics[width=\linewidth]{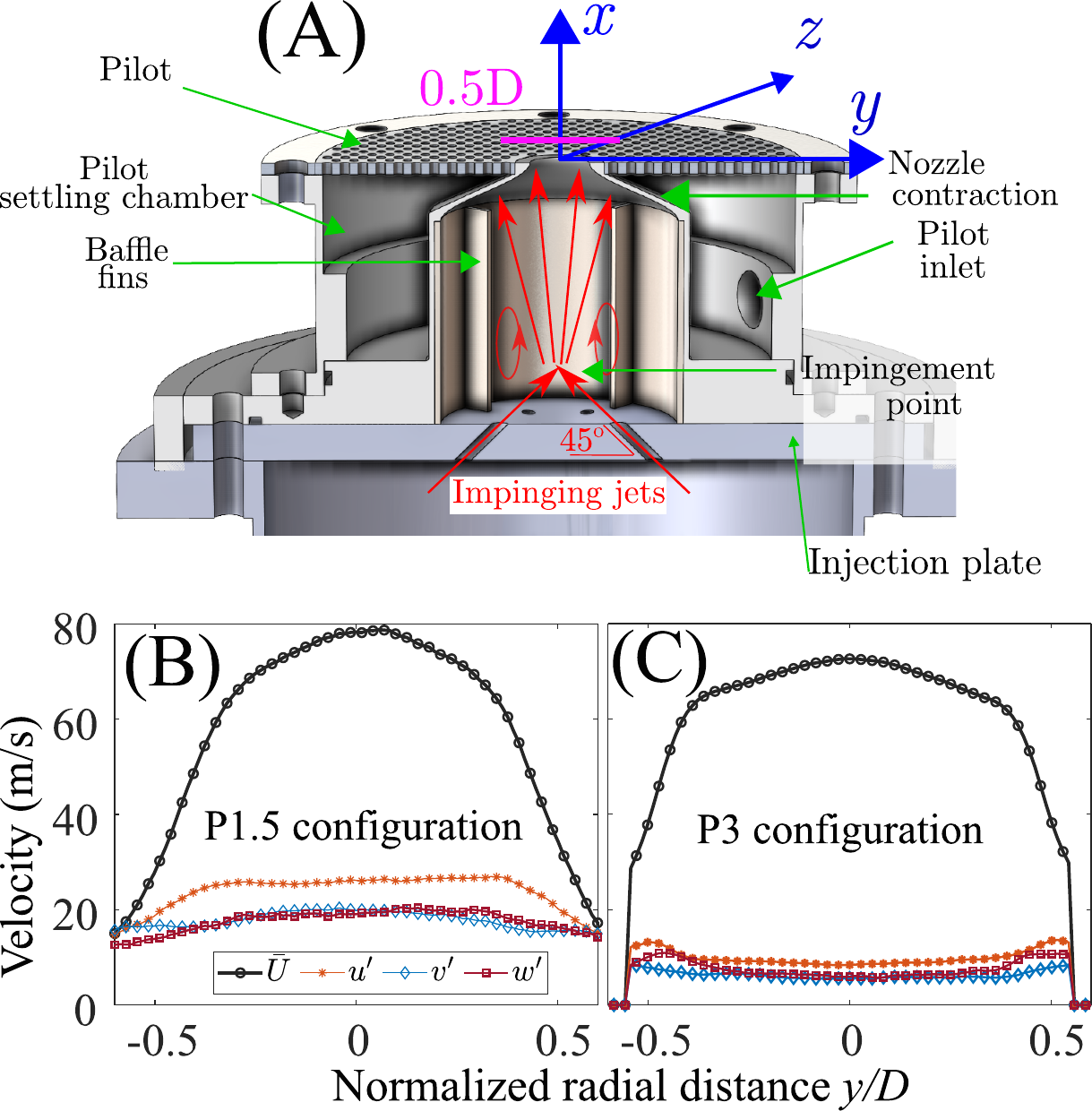} 
		\caption{\footnotesize Schematic of the burner setup (A) highlighting the turbulence generation mechanism and the radial variation of mean axial velocity and the rms fluctuating velocity components of the P1.5 (B) and P3 (C) configurations. The coordinate system used in measurements is marked for reference (A). The shown velocity profiles correspond to 60 m/s of bulk velocity and were taken 0.5$D$ downstream of the burner nozzle.}
		\label{schematic}
		\vspace{-2mm}
	\end{figure}
	
	The burner used in the study is shown in Fig. \ref{schematic}A and its turbulence characteristics have been described in \cite{kumar2025characteristics}. The high-turbulence P1.5 configuration and the low-turbulence P3 configuration of the burner are used, which comprise five 1.5 mm and six 3 mm tapered jets respectively, impinging upstream of the nozzle \cite{kumar2025characteristics}. The burner nozzle is 10 mm in diameter and is supported by a laminar pilot flame extending to 80 mm diameter. A perforated stainless-steel mesh is used to anchor the laminar pilot flame and sits flush with the nozzle exit. Reactants were metered and premixed upstream of the main and pilot settling chambers. Grade 4.0 purity CH$_4$ (99.99\%) was used for the studied main jet flame while the pilot flame was established using Grade 2.5 (99.5\%) CH$_4$. Filtered room temperature air (300 K) supplied by a central compressor facility was used as oxidizer for the main and pilot flames.
	
	The flow field at the burner exit was characterized using stereoscopic particle image velocimetry measurements of the central jet under non-reacting conditions. Room temperature (300 K) air was used as the working fluid and the pilot flame was not used. A dual-pulsed high speed Nd:YAG laser (Mesa PIV) was used to generate a 532 nm laser sheet of height 40 mm and FWHM of 500 $\mu$m at the burner axis. The laser was operated at 10 kHz and the flow was seeded with 0.2 $\mu$m $\mathrm{TiO}_2$ particles to ensure accurate flow tracing at elevated $u'$. The resulting scattering signals were collected using two high-speed cameras (Photron SA-5) arranged in a forward-scatter configuration. The cameras were fitted with a pair of telephoto lenses ( 200 mm f$\#$4) using Scheimpflug adapters and operated at a resolution of 896 x 848 pixels, allowing a field of view of 30 mm along the burner axis. A total of 3750 image pairs were collected for each condition and processed using the standard cross-correlation algorithm in LaVision DaVis10 software. Measurements were resolved using 16 x 16 pixel interrogation windows with 75\% overlap between vectors, corresponding to the final vector resolution of 0.56 mm with an adjacent vector spacing of 0.14 mm. The radial variations of mean and rms fluctuating velocity components for 60 m/s of bulk velocity are shown in Fig. \ref{schematic}. These cold flow measurements have been compared to reacting conditions (see Section 1 in Supplementary Materials) and remain valid.

	\begin{table*}[ht]
		\centering
		\small
		\caption{\footnotesize{Range of test conditions for CH$_4$--air mixtures.
			Turbulence quantities measured $0.5D$ downstream of the nozzle and spatially averaged over
			the central 50\% of the burner diameter are reported. The turbulent Reynolds numbers are defined as $\mathrm{Re}_{\mathrm{T}} = u'\Lambda/\nu_{\mathrm{air},300\,\mathrm{K}}$ and $\mathrm{Re}_{\mathrm{\lambda}} = u'_\lambda\lambda/\nu_{\mathrm{air},300\,\mathrm{K}}$. The Taylor and Kolmogorov length scales are estimated as $\lambda = \Lambda/\mathrm{Re}_{\mathrm{T}}^{1/2}$ and $\eta = \Lambda/\mathrm{Re}_{\mathrm{T}}^{3/4}$ and $u'_\lambda = u'(\lambda/\Lambda)^{1/3}$ \cite{DRISCOLL2020}. The Karlovitz and Damk\"ohler numbers are defined as $\mathrm{Ka} = (u'/S_{\mathrm{L}})^{3/2}(\Lambda/\delta_{\mathrm{L}})^{-1/2}$ and $\mathrm{Da} = \Lambda S_{\mathrm{L}}/(u'\delta_{\mathrm{L}})$, respectively. These quantities are reported for $\phi = 0.7$ and $1.0$.}}
		\label{table:test_conditions}
		
		\resizebox{\textwidth}{!}{%
			\begin{tabular}{c|c|cccccccccccccc}
				\hline
				        \rule{0pt}{5ex}
				 
				Case &
				\shortstack{$U_{\mathrm{bulk}}$\\(m/s)} &
				$\phi$ &
				\shortstack{$u'$\\(m/s)} &
				\shortstack{$v'$\\(m/s)} &
				\shortstack{$w'$\\(m/s)} &
				\shortstack{$\Lambda$\\(mm)} &
				\shortstack{$\lambda$\\($\mu$m)} &
				\shortstack{$\eta$\\($\mu$m)} &
				{$\mathrm{Re}$} &
				$\mathrm{Re}_{\mathrm{T}}$ &
				$\mathrm{Re}_{\lambda}$ &
				$\mathrm{Ka}_{0.7}$ &
				$\mathrm{Da}_{0.7}$ &
				$\mathrm{Ka}_{1.0}$ &
				$\mathrm{Da}_{1.0}$ \\
				\hline
				
				\multirow{7}{*}{P1.5}
				& 10 & 0.7, 0.8, 0.9, 1.0
				& 5.1 & 3.9 & 4.2 & 2.5 & 92 & 16
				& 6400 & 790 & 9.3 & 70 & 0.14 & 20 & 0.41 \\
				
				& 20 & 0.7, 0.8, 0.9, 1.0
				& 10.4 & 8.1 & 8.8 & 2.5 & 65 & 10
				& 12700 & 1640 & 11.8 & 200 & 0.07 & 60 & 0.20 \\
				
				& 40 & 0.7, 0.8, 0.9, 1.0
				& 20.8 & 16.0 & 17.8 & 2.5 & 48 & 6
				& 25500 & 3300 & 14.9 & 570 & 0.04 & 160 & 0.10 \\
				
				& 50 & 0.7, 0.8, 0.9, 1.0
				& 22.7 & 17.2 & 18.2 & 3.1 & 48 & 6
				& 31900 & 4480 & 16.5 & 590 & 0.04 & 170 & 0.11 \\
				
				& 60 & 0.7, 0.8, 0.9, 1.0
				& 26.2 & 19.7 & 20.3 & 3.2 & 44 & 5
				& 38200 & 5290 & 17.4 & 720 & 0.04 & 210 & 0.10 \\
				
				& 70 & 0.7, 0.8, 0.9, 1.0
				& 31.3 & 23.1 & 23.8 & 3.1 & 40 & 4
				& 44600 & 6150 & 18.3 & 950 & 0.03 & 280 & 0.10 \\
				
				& 80 & 0.8, 0.9, 1.0
				& 36.0 & 25.6 & 26.6 & 3.2 & 38 & 4
				& 51000 & 7250 & 19.4 & -- & -- & 340 & 0.08 \\
				
				\hline
				
				\multirow{5}{*}{P3}
				& 10 & 0.7, 0.8, 0.9, 1.0
				& 1.6 & 1.2 & 1.4 & 3.0 & 172 & 41
				& 6400 & 300 & 6.7 & 11 & 0.55 & 3 & 1.61 \\
				
				& 20 & 0.7, 0.8, 0.9, 1.0
				& 3.0 & 2.1 & 2.5 & 2.9 & 123 & 25
				& 12700 & 560 & 8.2 & 29 & 0.28 & 9 & 0.82 \\
				
				& 40 & 0.7, 0.8, 0.9, 1.0
				& 6.0 & 3.9 & 4.6 & 3.4 & 95 & 16
				& 25500 & 1300 & 10.9 & 75 & 0.17 & 23 & 0.48 \\
				
				& 50 & 0.7, 0.8, 0.9, 1.0
				& 7.3 & 5.0 & 5.9 & 3.6 & 88 & 14
				& 31900 & 1680 & 11.9 & 98 & 0.15 & 30 & 0.42 \\
				
				& 60 & 0.7, 0.8, 0.9, 1.0
				& 8.9 & 5.9 & 6.7 & 3.7 & 80 & 12
				& 38200 & 2070 & 12.7 & 131 & 0.12 & 40 & 0.35 \\
				
				\hline
			\end{tabular}%
		}
	\end{table*}

	PLIF measurements of OH and CH radicals were performed to assess flame speeds and reaction zone structure, respectively, and the methodology presented in \cite{carter2016high,mitsingas2019simultaneous} was adapted. Measurements were made using a high-speed, frequency-doubled dye laser (Sirah Credo) circulated with a DCM-ethanol solution. The dye was excited at 10 kHz using an Nd:YAG pump laser (Edgewave Innoslab). The resulting pulse was frequency doubled to excite the OH Q$_1$(8) $\textbf{A-X}$ transition at 309.24 nm. The pulse was focused into a laser sheet of height 70 mm and FWHM 110 $\mu$m (estimated by scanning a knife edge across the beam width) at the burner axis and had an average power of 3 Watts. The resulting fluorescence was collected using a high speed camera (Photron SA-Z) fitted with an image intensifier (UVI-1850) and a macro lens (Cerco-UV 94 mm F4.1). A narrow bandpass filter (Edmund Optics 325 $\pm$ 12.5 nm) was used to collect the red-shifted fluorescence from the same excitation band. A total of 4752 images with a resolution of 896x896 pixels were collected for each condition. The signal-to-noise ratio (SNR) of OH measurements, defined as the ratio of mean to standard deviation of signal in a 1 mm x 1 mm section of burnt products, was lowest for $\phi$ = 0.7 flames at 4.0-4.5 for raw images and improved to $>$ 5.0 by a 3 x 3 median filter. The mean signal strength across the flame front is higher than this baseline background signal and the presented statistics are not sensitive to the measured SNR and the filter kernel dimensions. The segmentation procedure yields converged statistics as verified by bootstrapping and comparison across independent measurements. CH-PLIF measurements were made using the same setup by exciting the Q$_1(7) \textbf{C-X}$(0,0) transition at 314.38 nm using the resonant detection methodology \cite{carter2016high}. The imaging optics were replaced with a Nikon UV 105 mm f4.5 lens. The laser sheet was collimated to a height of 45 mm and an imaging field of view of 37.5 mm x 37.5 mm was used to improve collection efficiency. A total of 4752 images with a resolution of 896 x 896 pixels were collected for each considered condition. OH- and CH-PLIF movies from selected cases have been included in the Supplementary Data for reference.
	
	\subsection{Data processing}
	\begin{figure}[h!]
		\centering
		\includegraphics[width=0.8\linewidth]{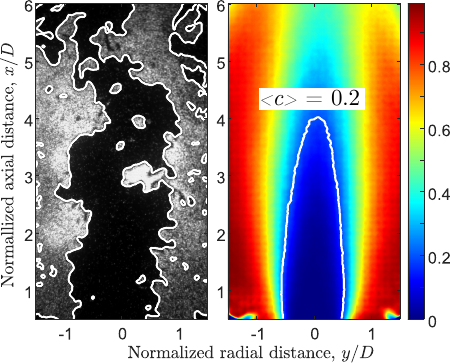}
		\caption{\footnotesize Instantaneous OH-PLIF image with marked flame boundary (left) and the ensemble-averaged flame brush with an overlaid iso-contour of the mean progress variable, $\left<c\right>$ = 0.2 considered in analysis. Data for the $U_\mathrm{bulk}$ = 60 m/s and $\phi$ = 0.7 condition of the P1.5 configuration is shown.}
		\label{data_processing}
		\vspace{-2mm}
	\end{figure}

	The relevant turbulence statistics are calculated from the resolved velocity vectors. Bulk velocity, $U_\mathrm{bulk}$, is defined as the volumetric flow rate divided by the nozzle area, $\dot{m}/\rho_\mathrm{o} A_{\mathrm{nozzle}}$, with $\rho_\mathrm{o}$ evaluated at the nozzle exit. Turbulence parameters 0.5$D$ downstream of the nozzle are reported and averaged over the central 50\% of the nozzle, with integral length scale, $\Lambda$, estimated using the spatial auto-correlation of fluctuating velocity:
	
	\begin{equation}
		\Lambda = \sum_{n = 0}^{N}\frac{\langle {u''(x)u''(x+ n\Delta h)}\rangle}{{{u'(x)}} {u'(x+ n\Delta h)}} \Delta h
	\end{equation}
	where $u''(x)$ and $u'(x)$ are the instantaneous and rms fluctuating velocity at location $x$, $\langle \rangle$ denotes ensemble averaging and $\Delta h$ denotes the spacing between adjacent PIV velocity vectors. The summation of the auto-correlation function was performed up to its first zero crossing, \textit{N}. The Taylor and Kolmogorov scales are estimated from the scaling relations, $\lambda = \Lambda/Re_{\mathrm{T}}^{0.5}$ and $\eta = \Lambda/Re_{\mathrm{T}}^{0.75}$ respectively, where $\mathrm{Re}_\mathrm{T}$ is the turbulent Reynolds number \cite{pope_2000}. The Karlovitz number, Ka is taken to be $\left(u'/S_\mathrm{L}\right)^{1.5}\left(\Lambda/\delta_\mathrm{Lo}\right)^{-0.5}$ \cite{DRISCOLL2020}.
	
	Several methods of estimating flame reaction rates have been suggested \cite{DRISCOLL200891} and can vary in estimates and scaling of the reaction rates. While flame tracking methods are essential for transient \cite{bradley2020flame} and open \cite{francolini2024investigation} flame configurations, enveloped flames present a relatively straightforward means to estimate flame speeds. Global consumption flame speed, $S_{\mathrm{T,GC}}$, is defined as the average velocity of the flame front required to consume the supplied volume of reactants within a given area. This area is tracked using suitable reaction markers. In the present study, flame speeds are estimated from the binarized OH-PLIF images. Images were background subtracted and median filtered prior to intensity normalization. Images were binarized using Otsu's segmentation \cite{otsu1979threshold} and the reference codes provided in \cite{chaib2023hybrid} were adapted. These binarized OH-PLIF images were averaged to generate a mean flame brush encoding the statistical location of the flame front, as shown in Fig. \ref{data_processing}. Further samples from $\phi$ = 0.7 and 1.0 conditions have been included in section 5 of Supplementary Materials. The magnitude of the ensemble-averaged flame brush at a given location denotes the proportion of the deficient reactant consumed within its subtended volume and is referred to as the mean progress variable, $\langle c \rangle$. The envelope subtended by the flame brush for a given $\langle c \rangle$ is then estimated by integrating the revolved surface of its two-dimensional iso-contour. Finally, global consumption flame speeds are estimated using the relation \cite{DRISCOLL200891}: \vspace{-2mm}
	\begin{equation}
		\label{global_consumption_speed}
		\dot{m} = \rho_\mathrm{o} S_{\mathrm{T,GC, \langle c \rangle}} A_\mathrm{\left<c\right>}  \vspace{-2mm}
	\end{equation} 
	In the present study, measurements were made using the $\left<c\right> = 0.2$ iso-contour of the mean progress variable as shown overlaid on the ensemble-averaged flame brush in Fig. \ref{data_processing}.  
	
	\subsection{Uncertainty analysis}
	Errors in the estimation of flame speeds are quantified from the area-velocity relation given by Eq. \ref{global_consumption_speed}. Uncertainties in the reactant mass flow rates derive from flow metering errors and are taken to be $\pm$1\% of the full-scale range of the flow controllers. This error is proportionally weighted for individual reactant streams, resulting in a total uncertainty of 2.9\% at 10 m/s of bulk velocity and 0.9\% at 80 m/s. A constant 1\% variation in density effects is assumed due to temperature variations. The uncertainties associated with mean and fluctuating velocity calculations are within 1\% of their corresponding reported values.
	
	The resolution of the OH-PLIF imaging system is quantified using a USAF target plate to be 140 $\mu$m based on a 20\% cutoff of the modulation transfer function and is less than the respective laminar flame thickness of the studied flames. The uncertainty in flame area estimation is limited by the segmentation accuracy, which depends on the flame thickness, $\delta_\mathrm{l}$, and is taken to be $\pm$$P\delta_\mathrm{l}/2$, where $P \propto \sqrt{A_\mathrm{T}}$ is the perimeter of the flame. This uncertainty reduces from 1.1\% for 10 m/s of bulk velocity to 0.6\% at 80 m/s. The variance in the measured $A_\mathrm{T}$ is verified for convergence by the bootstrapping method and is within 1.0\%. Similarly, the errors associated with the binarization process are estimated at $\pm6\%$ (see Section 3 in Supplementary Materials). Combining the individual uncertainty contributions, the measured flame speeds have uncertainty bounds of $\pm$ 7\% and are shown overlaid on Fig. \ref{St_Sl0} and Fig. \ref{St_SlP3} for reference.
	
	\subsection{Test conditions}  
	\begin{figure}[h!]
		\centering
		\includegraphics[width=0.7\linewidth]{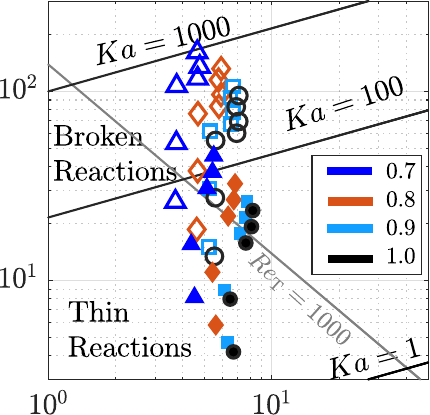}
		\caption{\footnotesize Experimental test conditions realized in the present study overlaid on the regime diagram \cite{Peters_2000}. Colors denote the equivalence ratio, $\phi$, while the open and closed symbols represent the P1.5 and P3 configurations, respectively.}
		\label{regime_diagram}
		\vspace{-2mm}
	\end{figure}
	
	Premixed CH$_4$-air flames with equivalence ratios ranging from 0.7-1.0 at bulk velocities up to 80 m/s are studied. For the P1.5 configuration, measurements at $\phi$ = 0.7 were limited to 70 m/s of bulk velocity due to limited pilot backsupport. Sample binarized images from representative test conditions have been provided in Section 5 of Supplementary Materials for comparison. The variation of upstream turbulence with mixture composition is assumed limited (see Supplementary Materials Section 1) and the same turbulence parameters are shared across the considered mixture compositions. These are compared against flame speed measurements of the P3 configuration of the burner which generates moderate turbulence intensities (15\% of bulk velocity). The pilot flame was stabilized at a volumetric flow rate of 263 SLPM and was matched to the respective jet flame equivalence ratio for all considered cases. The pilot thermal power therefore varies with equivalence ratio and the output power scales from 9.7-13.5 kW for $\phi$ = 0.7-1.0. At $\phi$ = 0.7, the pilot flow rate was increased to 322 SLPM for bulk velocities exceeding 50 m/s in the P1.5 configuration (maximum setpoint on oxidizer flow controller) to limit downstream cold air entrainment, increasing the total thermal power to 11.9 kW for these cases. Limited interaction between the first row of pilot flamelets and the main jet is noted for the $\phi$ = 0.7 flames as some mesh orifices were closer to the main jet (see section 4 in Supplementary Materials). These are not expected to affect the measured statistics. Similarly, beyond the highest considered bulk velocity for $\phi$ = 0.7 flames, air entrainment effects become severe. The use of upstream $\langle c \rangle$ = 0.2 progress variable is partially motivated to avoid potential entrainment artifacts. Note that the limited downstream backsupport and air entrainment do not affect the statistics derived from fully supported upstream regions since the relevant local boundary conditions remain unchanged (see Section 6 of Supplementary Materials). The reference laminar flame properties were estimated using freely propagating laminar flame simulations \cite{cantera} using the GRI 3.0 mechanism \cite{grimech}. The normalized turbulence parameters are overlaid on the regime diagram \cite{Peters_2000} and span from thin-reaction zone to the broken reaction zone regime as shown in Fig. \ref{regime_diagram}. The final test conditions and relevant turbulence parameters are summarized in Table \ref{table:test_conditions}. The measured test conditions denote a progressively increasing proportion of TKE in the smaller scales of turbulence with limited variability in the larger integral length scales.


		\section{Results}
			\label{section3}
		\subsection{Flame speed measurements}

		\begin{figure}[h!]
			\centering
			\includegraphics[width=0.8\linewidth]{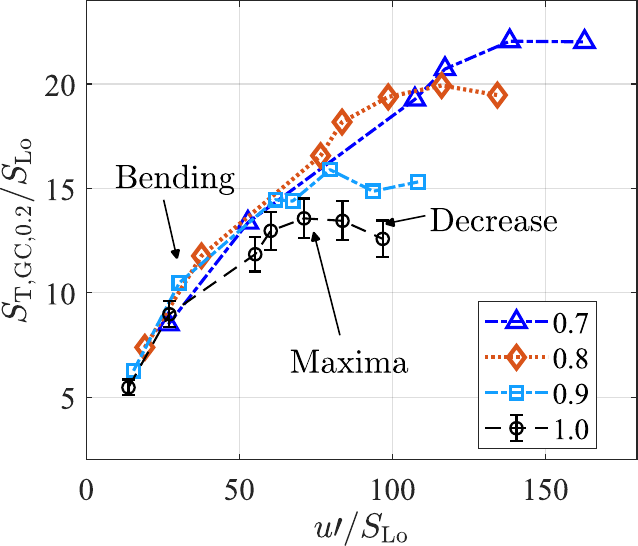}
			\caption{\footnotesize Variation of the normalized global flame speed, $S_\mathrm{T,GC,0.2}$, based on the 0.2 iso-contour of mean progress variable, with normalized $u'$ at the burner exit for the P1.5 configuration.}
			\label{St_Sl0}
			\vspace{-2mm}
		\end{figure}

		\begin{figure}[h!]
			\centering
			\includegraphics[width=\linewidth]{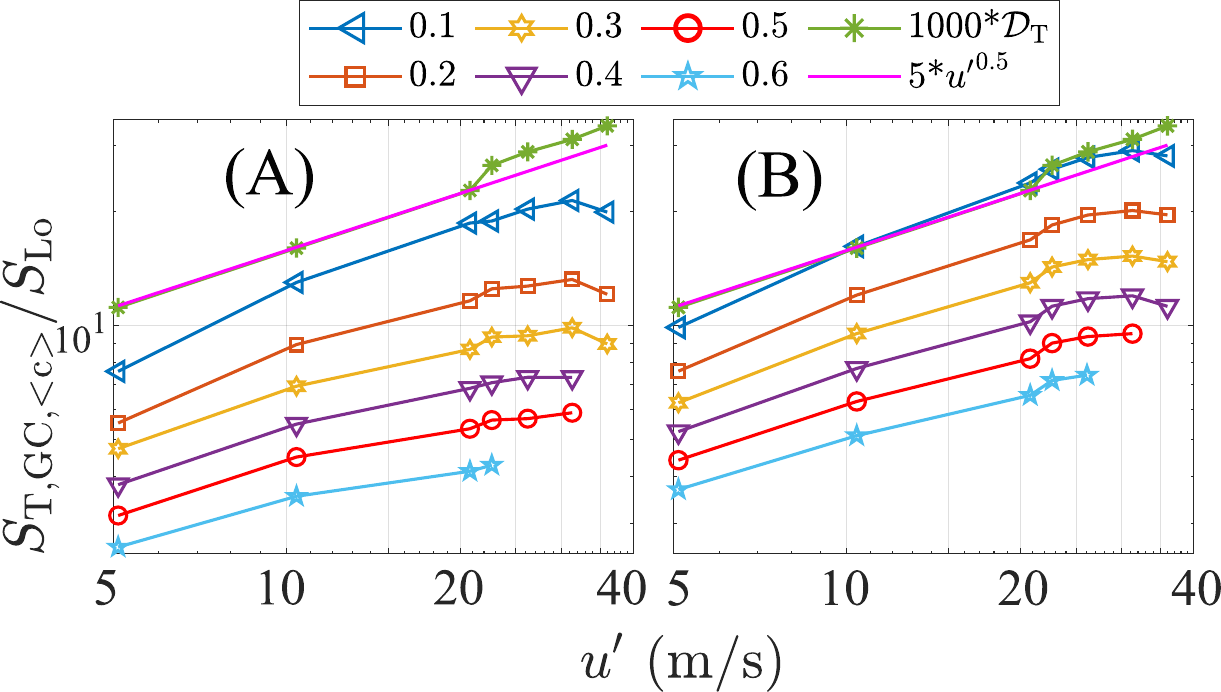}
			\caption{\footnotesize Variation of the normalized global consumption flame speed, $S_\mathrm{T,GC,\left<c\right>}$, derived from the iso-contour  $\left<c\right>$ of the mean progress variable, with the mean axial turbulence at the burner exit. Panels (A) and (B) correspond to $\phi$ = 1 and $\phi$ = 0.8 test conditions of the P1.5 configuration respectively. Also overlaid are the $\mathcal{D}_\mathrm{T}$ and $u'^{0.5}$ scalings for the corresponding test conditions.}
			\label{isocontours}
						\vspace{-2mm}
		\end{figure}
		\begin{figure*}[h!]
			\centering
			\includegraphics[width=0.65\linewidth]{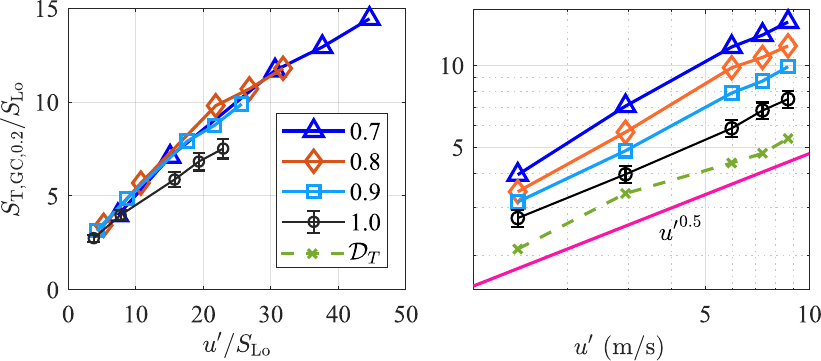}
			\caption{\footnotesize  Variation of the normalized global flame speed, $S_\mathrm{T,GC,0.2}$, based on the 0.2 iso-contour of mean progress variable, with normalized (left) and absolute (right) rms fluctuating velocity at the burner exit for the P3 configuration.} 
			\label{St_SlP3}
			\vspace{-2mm}
		\end{figure*}
		The response of premixed flames to upstream turbulence is estimated using the measured turbulent flame speeds. These measurements are scaled to their respective freely propagating laminar flame speeds. Flame speeds increase with turbulence and show the characteristic bending behavior reported in past studies \cite{gulder2007fsd,wabel2017turbulent,ahmed2019insights}, see Fig. \ref{St_Sl0}. However, as the turbulence is increased further, the flame speeds approach a maximum and then appear to decrease. This is in contrast with the established ${\mathcal{D}_\mathrm{T}}$ scaling reported in literature and suggests that turbulence does not enhance reaction rates indefinitely. The observed decrease, however, is not due to the entrainment of cold ambient air since sufficient pilot backsupport was ensured. Similar stalls in flame speeds have been reported in numerical studies of high Karlovitz number premixed flames \cite{aspden_day_bell_2019,russell2025turbulence} but were not realized in experimental studies \cite{wabel2017turbulent,wang2019structure} despite access to higher $u'$ compared to the present study. We note that, despite similar $u'$, proportionally higher Karlovitz numbers are realized in the current study, with a larger fraction of TKE residing at timescales exceeding flame timescales.
		
		Observing the variation of flame speeds between the equivalence ratios, leaner flames burn significantly faster compared to stoichiometric mixtures when scaled with their laminar flame speeds. We report a similar trend in the flame speed measurements of C$_3$H$_8$-air mixtures realized using the P1.5 configuration, with leaner flames experiencing proportionally greater flame speed enhancements (see Section 2 in Supplementary Materials). However, unlike the CH$_4$-air flames, C$_3$H$_8$-air flame speeds scale monotonically in the measured range of test conditions with no observed burning limits. The laminar flame speeds and adiabatic flame temperatures of the two fuels are comparable. However, the Zeldovich flame thicknesses of C$_3$H$_8$ mixtures are less than 50\% of the corresponding CH$_4$-air mixtures \cite{tamadonfar2015effects}, limiting the flame's access to TKE at the relevant small scales. The limits of burning for C$_3$H$_8$-air mixtures thus exceed those of CH$_4$-air flames.
		
		The 0.2 iso-contour of the mean progress variable was chosen to limit the downstream entrainment effects present at large $u'$. To further ascertain that the observed trends do not depend on the choice of iso-contour, the variation of global flame speeds with the measured upstream $u'$ is compared across the iso-contours of the mean progress variable in Fig. \ref{isocontours}. Also shown is the evolution of $\mathcal{D}_\mathrm{T}$ and $\sqrt{{u'}}$ to highlight the DH2 scaling \cite{damkohler1947effect}. Measurements based on the 0.6 iso-contour were limited to bulk velocities of 50 m/s and 60 m/s for $\phi$ = 1.0 and 0.8 respectively, beyond which the iso-contours exceeded the measurement domain. Since the larger progress variables subtend a greater area (see Fig. \ref{data_processing}), the magnitude of the measured flame speed decreases for the same mass flow rate of reactants as given by Eq. \ref{global_consumption_speed}. The measured flame speeds scale similarly and the observed flame speed maxima occur at approximately the same $u'$ across the considered mean progress variable iso-contours. Furthermore, the flame speeds scale with turbulent diffusivity up to $u'$ = 26.2 m/s ($U_\mathrm{bulk}$ = 60 m/s) and then diverge for both equivalence ratios considered in the analysis despite having different Karlovitz numbers. In contrast, the $\mathcal{D}_\mathrm{T}$ scaling extends well beyond the current measurements for piloted jet flames \cite{wang2019structure}.
		
		Variations in the normalized global consumption speeds with normalized $u'$ for the low turbulence intensity P3 configuration are shown in Fig. \ref{St_SlP3}A. The measurements exhibit a monotonic increase with $u'$, with magnitudes comparable to equivalent measurements reported for the P1.5 configuration (Fig. \ref{St_Sl0}). Similarly, as observed for the P1.5 configuration, stoichiometric flames show limited sensitivity to $u'$ compared to leaner mixtures. To facilitate comparison between equivalence ratios, the normalized flame speeds are compared against $u'$ in Fig. \ref{St_SlP3}B. The measured flame speeds exhibit a weak initial divergence and then scale with $\mathcal{D}_\mathrm{T}$. At low $u'$, flame speed enhancement is governed by flame wrinkling \cite{damkohler1947effect}, wherein turbulence generates wrinkling most effectively for the leanest (highest Ka) flames \cite{tamadonfar2015effects}, resulting in the initial divergence between the flame speeds. However, once large scale effects saturate, a consistent scaling across equivalence ratios corresponding to small scale effects is realized. This transition approximately coincides with the third test condition ($U_\mathrm{bulk}$ = 40 m/s, $u'$ = 6.0 m/s) of the P3 configuration across all equivalence ratios, despite substantial differences in their Karlovitz numbers (23 and 75 at $\phi$ = 1.0 and 0.7, respectively). This suggests that the conventional scaling parameters derived using laminar flames remain insufficient to describe such transitions \cite{wang2025experimental,chaib2026normalization}. Beyond this transition, the divergence established by large-scale flame wrinkling persists, while further enhancements in flame speed arise from local mixing effects superimposed on the existing contributions. In addition to turbulence, these large-scale effects depend on the bulk velocity and burner geometry \cite{filatyev2005}, and their quantification is beyond the scope of the present study. Note that the measurements obtained using the P1.5 configuration extend beyond this transition and conform to the $\mathcal{D}_\mathrm{T}$ scaling in its initial range. 
	
		\begin{figure}[h!]
			\centering
			\includegraphics[width=0.7\linewidth]{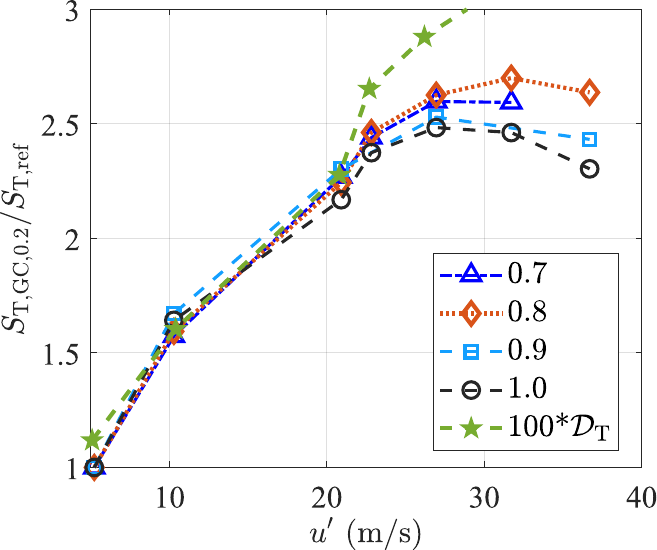}
			\caption{\footnotesize Variation of the global flame speed, $S_\mathrm{T,GC,0.2}$, based on the 0.2 iso-contour of mean progress variable, with the rms fluctuating velocity at the burner exit. $S_{\mathrm{T,ref}}$ is the measured $S_\mathrm{T,GC,0.2}$ at $U_\mathrm{bulk}$ = 10 m/s.}
			\label{St_Slref}
		\vspace{-2mm}
		\end{figure}
		\begin{figure*}[h!]
			\centering
			\includegraphics[width=0.6\textwidth]{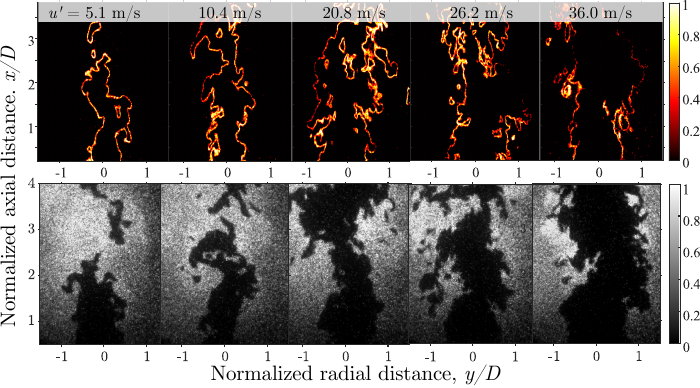}
			\caption{\footnotesize Instantaneous CH-PLIF (top) and OH-PLIF (bottom) images from the P1.5 configuration. Images correspond to $\phi$ = 1.0 at increasing $u'$.}
			\label{CH_images}
			\vspace{-2mm}
		\end{figure*}
		
		The presented flame speed measurements at large $u'$ (Fig. \ref{St_Sl0}) display significant scatter, with leaner flames showing proportionally larger enhancements in flame speeds compared to their unstretched laminar counterparts. As discussed, this suggests that the chosen basis for normalization is not adequate. A similar trend is realized in the turbulent jet flame studies \cite{zhou2015distributed}, wherein the authors normalized the measured flame speeds with corresponding measurements of a low speed jet in the same setup. It should be noted that their measured laminar flame speeds for $\phi$=0.7 flames were as large as unstretched $\phi$=0.9 flames (increase of 73\%) while $\phi$=1.0 flames were only enhanced by 20\%. While the noted variations in the mentioned study were attributed to the use of a higher equivalence ratio pilot, similar trends are evident in the current measurements despite pilot matching. Once these initial variations are accounted for, a consistent scaling reflecting the role of small scales of turbulence can be recovered. The flame speeds for each of the studied equivalence ratios are thus normalized by their respective measured flame speeds at $U_\mathrm{bulk}$ = 10 m/s, hereafter referred to as $S_{\mathrm{T,ref}}$. The resulting normalized flame speeds are independent of the reactant equivalence ratio and show a consistent scaling with the measured upstream turbulence, see Fig. \ref{St_Slref}. The measured flame speeds exhibit maxima at approximately the same $u\prime$ despite the prescribed variations in the respective mixture and ambient pilot compositions, suggesting that turbulent mixing effects dictate the observed scaling over variations in mixture equivalence ratio in high Karlovitz number regimes. This maximum occurs at $u'$ = 31.3 m/s and Re$_\mathrm{T}$ = 6150 corresponding to a bulk velocity of 70 m/s. A similar limit for the broadening of preheat layers is proposed at Re$_\mathrm{T}$ = 2800 or Re$_\lambda$ = 13.8 \cite{DRISCOLL2020,skiba2018part1} by comparing the turbulence diffusivity of smaller scales to molecular diffusivity at the relevant flame scales (CH$_2$O layer thickness). As expected, the measured flame speed limits exist beyond the preheat layer broadening limit. Despite normalization, measurements exhibit some residual scatter which becomes apparent around the maxima. This correlates with the limited but consistent dissipation of the small scales of turbulence across the flame front, which tends to increase with flame temperature \cite{towery2016spectral}. Note that this scatter, although weak, is also observed in the reported measurements of jet flames \cite{wang2019structure}.		

		\subsection{Flame structure}

		To assess how the flame structure evolves with turbulence, qualitative assessments of the reaction zone structure were made using CH-PLIF. Instantaneous CH-PLIF images of the flame front with varying rms fluctuating velocity are shown in Fig. \ref{CH_images}. Also shown are the corresponding OH-PLIF snapshots to confirm the extent of backsupport in the considered field-of-view. At the low $u'$ of 5.1 m/s, a continuous flame front exists and the CH signals are limited to thin layers. With an increase in $u'$ up to 20.8 m/s, flames exhibit a consistent increase in wrinkling and local straining, but the measured flame front remains intact and does not exhibit any large-scale extinctions. This observation is consistent with the thin reaction zones observed in the Hi-Pilot geometry \cite{skiba2018part1}. A further increase in $u'$ leads to progressively larger areas of the flame front becoming strained and the flames exhibit extinction behavior. Note that locally, the flamelet-like structure \cite{DRISCOLL2020} remains preserved and no smearing or broadening of CH layers is observed. A similar picture was not confirmed for the Hi-Pilot geometry where the reaction layers were shown intact for $u'$ up to 39 m/s. In comparison, extinctions have been reported in backsupported counterflow flames \cite{coriton2013effects,tirunagari2017characterization} while broadened flame structures \cite{zhou2015distributed,zhou2017thinreaction,dunn2010finite} have been shown to exist in jet flames. Among these configurations, the reported Kolmogorov length scales are the smallest in the present study while the Hi-Pilot geometry \cite{skiba2018part1,wabel2017turbulent} reported the largest measured axial length scales. These large scales appear to preserve the underlying flame structure \cite{aspden2010distributed,skiba2018part1} while the smaller scales tend to modify or disrupt it \cite{peters_1999}. Similarly, the upstream reactant preheating effects are expected to be the largest in jet flames \cite{chen1996detailed}. The role of preheating thus appears consistent with the existence of broadened structures as concluded in \cite{DRISCOLL2020}, with an analogous turbulent mixing with cold reactants leading to extinctions.
		
		\subsection{Modified flame speed scaling}
\begin{figure}[h!]
			\centering
			\includegraphics[width=0.75\linewidth]{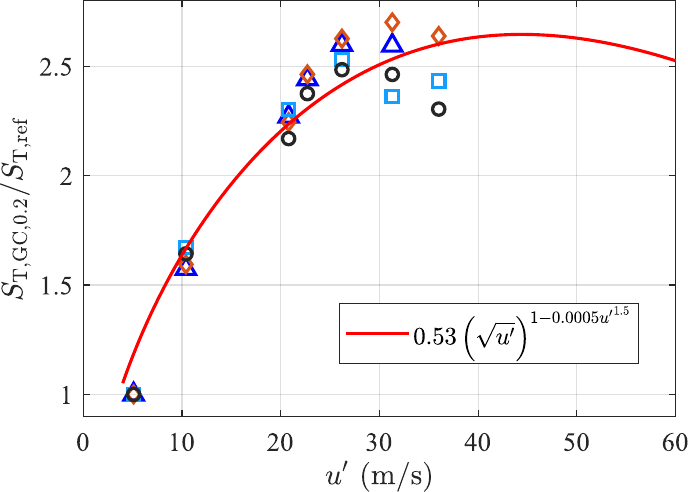}
			\caption{\footnotesize Proposed flame speed scaling compared with the experimental measurements.}
			\label{St_scaling}
			\vspace{-5mm}
		\end{figure}

		The observations of premixed flame speed and flame structure evolution with turbulence discussed in the preceding sections can be summarized as follows: At low $u'$, the burning rates of premixed flames are enhanced by large scale wrinkling. Beyond a certain threshold, these wrinkling effects saturate and the burning rates instead scale with turbulent diffusivity, $\mathcal{D}_\mathrm{T}$. As the proportion of TKE in the small scales increases, flames become increasingly strained and the measured flame speeds exhibit a departure from the $\mathcal{D}_\mathrm{T}$ scaling. A critical threshold is then reached, at which the flame speed enhancement by large scales is balanced by the progressively increasing contributions from the small scales towards straining and extinction events. Beyond this threshold, burning rates are limited by extinctions and the measured flame speeds appear to stall or decrease. A similar transition occurs at larger scales \cite{aspden2010distributed}, where excess TKE in the Da$>$1 regime does not contribute to flame speeds. Note that at any given instant, some portions of the flame front will be excessively strained while other segments will experience enhanced reaction rates with the measured global flame speeds highlighting the average behavior. This competition of flame speed enhancement due to larger energy-containing scales and extinctions due to straining at the smaller scales in the DH2 regime can be modeled as a departure from the baseline $\mathrm{\mathcal{D}_T}$ scaling:
		\vspace{-2mm}
		\begin{equation}
			\label{scaling_eqn}
			S_\mathrm{T} \propto {(\mathcal{D}_\mathrm{T})}^{1-f(u')}
		\end{equation}
		where $f(u')$ is a function of $u'$ and quantifies extinction propensities. In spherical flame studies, extinction limits have been described using critical strain rates \cite{bradley2020flame,KIDO20021855}, wherein the flame kernels fail to propagate when the Karlovitz stretch factor, ${K}$, defined as the ratio of chemical time scales ($\delta_\mathrm{L}/S_\mathrm{L}$) to flow timescales based on Taylor length scale ($\lambda/u'$), exceeds unity. Note that both the stretch factor, $K$ and Karlovitz number, Ka, denote the same scaling of the turbulence dissipation rate, ${\varepsilon}$, and can be used interchangeably. However, the selection of reference flame properties remains critical. In spherical flames, these reference properties are estimated from the initial flame kernel or using Markstein lengths \cite{KIDO20021855,chaudhuri2013scaling}. As discussed in Section 3.1, predicting these values for piloted flames remains challenging. Instead, the flame speed measured under the lowest $u'$ condition of the P1.5 configuration is taken as the reference flame speed. We assume $f(u') \propto {K}$ as a first approximation. The proposed scaling can thus be represented as:
		\vspace{-2mm}
		\begin{equation}
			\label{scaling_general}
			({S_\mathrm{T}}/{S_\mathrm{T,ref}}) = \alpha{(\mathcal{D}_\mathrm{T})}^{1-\beta{K}}
		\end{equation}
		where $\alpha$ and $\beta$ are assumed constant. Further, $K$ (or Ka) can be approximated as ${u'}^{1.5}\Lambda^{-0.5}$. For a constant $\Lambda$, Eq. \ref{scaling_general} reduces to an expression in $u'$. The measured flame speeds are then fitted to the proposed expression, yielding a scaling of the form:		\vspace{-0.5em}
		\begin{equation}
			\frac{S_\mathrm{T}}{S_\mathrm{T,ref}}=0.53\left(\sqrt{u'}\right)^{\,1-0.0005\left(u'^{1.5}\right)}
		\end{equation}
		and is presented alongside the experimental measurements in Fig. \ref{St_scaling}. Note that the fitted coefficient $\beta$ in the exponent is small, highlighting that energy at smaller scales must increase sufficiently before these effects become apparent. The expression can also be extended to accommodate the observed deviations in the $Da >$1 regime \cite{aspden2011turbulence}. This proposed scaling, however, is limited to the discussed measurements and the contributions of large scales were effectively accounted for through normalization by $S_\mathrm{T,ref}$. A more general formulation should build upon Eq. (\ref{scaling_general}) by incorporating the normalized contributions from larger scales and the intermittency of the small scales.
		
		\section{Summary and conclusions}
			\label{section4}
		The role of small scale turbulence on the burning rates of premixed flames was studied in a piloted high Karlovitz number burner. Premixed CH$_4$-air mixtures with $\phi$ = 0.7-1.0 were subjected to rms fluctuating velocities up to 36.0 m/s and a Karlovitz number of 950. The effects of turbulence were quantified using global consumption flame speeds, $S_\mathrm{T,GC}$, derived from OH-PLIF measurements while qualitative flame structure assessments were used to support the observations. The key findings of the study are summarized as follows:
		
		\begin{itemize}
			\item At low $u'$, turbulence enhances premixed flames through area generation with leaner mixtures exhibiting a greater sensitivity to $u'$ compared to stoichiometric flames. Once these effects saturate, a consistent $\mathcal{D}_\mathrm{T}$ scaling acts upon these saturated wrinkling effects across all equivalence ratios.
			\item The measured flame speeds increase with $u'$ up to a critical threshold, beyond which a further increase in $u'$ leads to progressively lower flame speeds. This critical threshold lies above the preheat layer broadening limit reported in the literature \cite{skiba2018part1}.
			\item The observed burning limits are shown to be independent of equivalence ratio when the measured flame speeds are scaled with $u'$.
			\item CH layers remain localized throughout the measurement domain. An increase in $u'$, however, does reveal progressive straining and local extinctions of the reaction zone. OH-PLIF imaging shows no cold air entrainment artifacts despite apparent gaps in CH signals.
			\item The observed flame speed scaling is modeled as a deviation from the baseline $\mathcal{D}_\mathrm{T}$ scaling due to extinctions at smaller scales. Extinctions are modeled using the Karlovitz stretch factor, $K$.    
		\end{itemize}
		
		\acknowledgement{CRediT authorship contribution statement} \addvspace{10pt}
		{\bf PK}: Conceptualization, Data curation, Formal analysis, Investigation, Methodology, Validation, Writing - original draft, Writing - review and editing. 
		
		{\bf\"{O}LG}: Conceptualization, Funding acquisition, Project administration, Resources, Supervision, Validation, Writing - review and editing.
		\acknowledgement{Declaration of competing interest} \addvspace{10pt}
		The authors declare that they have no known competing financial interests or personal relationships that could have appeared to influence the work reported in this paper.
		\acknowledgement{Acknowledgments} \addvspace{10pt}
		The authors acknowledge the financial support from the Natural Sciences and Engineering Council of Canada for the current work.	
		
		\footnotesize
		\baselineskip 9pt
		
		\bibliographystyle{proci}
		\bibliography{PROCI_LaTeX}
		
		
		\newpage
		
		\small
		\baselineskip 10pt
		
		
	\end{document}